\documentclass[twocolumn,showpacs, superscriptaddress,prl,amsmath,amssymb]{revtex4}

\usepackage{graphicx}
\usepackage{dcolumn}
\usepackage{bm}
\usepackage{epstopdf}

\begin{document}

\title{High pressure floating zone growth and structural properties of ferrimagnetic quantum paraelectric BaFe$_{12}$O$_{19}$ }

\author{H. B. Cao}
\affiliation{Quantum Condensed Matter Division, Oak Ridge National Laboratory, Oak Ridge, Tennessee 37831, USA}

\author{Z. Y. Zhao}
\affiliation{Materials Science and Technology Division, Oak Ridge National Laboratory, Oak Ridge, Tennessee 37831, USA}
\affiliation{Department of Physics and Astronomy, University of Tennessee, Knoxville, Tennessee 37996, USA}

\author{M.~Lee}
\affiliation{Department of Physics, Florida State University, Tallahassee, FL 32306-3016, USA}
\affiliation{National High Magnetic Field Laboratory, Florida State University, Tallahassee, FL 32310-3706, USA}

\author{E.~S.~Choi}
\affiliation{National High Magnetic Field Laboratory, Florida State University, Tallahassee, FL 32310-3706, USA}

\author{M. A. McGuire}
\affiliation{Materials Science and Technology Division, Oak Ridge National Laboratory, Oak Ridge, Tennessee 37831, USA}

\author{B. C. Sales}
\affiliation{Materials Science and Technology Division, Oak Ridge National Laboratory, Oak Ridge, Tennessee 37831, USA}

\author{H. D. Zhou}
\affiliation{Department of Physics and Astronomy, University of Tennessee, Knoxville, Tennessee 37996, USA}

\author{J.-Q. Yan}
\affiliation{Materials Science and Technology Division, Oak Ridge National Laboratory, Oak Ridge, Tennessee 37831, USA}
\affiliation{Department of Materials Science and Engineering, University of Tennessee, Knoxville, Tennessee 37996, USA}

\author{D.~G.~Mandrus}
\affiliation{Materials Science and Technology Division, Oak Ridge National Laboratory, Oak Ridge, Tennessee 37831, USA}
\affiliation{Department of Materials Science and Engineering, University of Tennessee, Knoxville, Tennessee 37996, USA}

\date{\today}

\begin{abstract}
High quality single crystals of BaFe$_{12}$O$_{19}$ were grown using the floating zone technique in flowing oxygen pressurized to 100 atm. Single crystal neutron diffraction was used to determine the nuclear and magnetic structure of BaFe$_{12}$O$_{19}$ at 4~K and 295~K. At both temperatures, there exist local electric dipoles formed by the off-mirror-plane displacements of magnetic Fe$^{3+}$ ions at the bipyramidal sites. The displacement at 4~K is about half of that at room temperature. The temperature dependence of the specific heat shows no anomaly associated with long range polar ordering in the temperature range from 1.90-300~K. The inverse dielectric permittivity, $1/\varepsilon$, along the $\emph{c}$-axis shows a $T^2$ temperature dependence between 10~K and 20~K, with a significantly reduced temperature dependence displayed below 10 K. Moreover, as the sample is cooled below 1.4~K there is an anomalous sharp upturn in $1/\varepsilon$. These features resemble those of classic quantum paraelectrics such as SrTiO$_3$. The presence of the upturn in $1/\varepsilon$ indicates that BaFe$_{12}$O$_{19}$ is a critical quantum paraelectric system with Fe$^{3+}$ ions involved in both magnetic and electric dipole formation.

\end{abstract}

\pacs{75.50.Gg; 
05.30.Rt; 
42.50. Lc;
77.84.-s;
61.05.fm;
75.85.+t; 
}

\maketitle

\section{Introduction}

Quantum paraelectricity has been an interesting topic for decades. A well-known case is SrTiO$_3$, which displays a very large dielectric constant at low temperatures but fails to go through a ferroelectric transition due to quantum fluctuations associated with the zero point energy \cite{srtio3}. The quantum paraelectric state is fragile and can be tuned to the ferroelectric state with a small stimulus such as quenched disorder, isotopic substitution, and external stress \cite{devonshire, bednorz, ang, itoh, haeni}. Quantum paraelectrics therefore provide unique examples to study quantum phase transitions that may underlie much universal behavior \cite{quantum_paraelectric}.

Combining magnetism with quantum paraelectricity is a topic of lively recent interest. EuTiO$_3$, with magnetic Eu$^{2+}$ ions replacing nonmagnetic Sr$^{2+}$, has attracted considerable attention as an antiferromagnetic quantum paraelectric system, in which magnetism can be employed as a parameter to tune the dielectric response. \cite{takagi} The delicate balance in EuTiO$_3$ between para- and ferroelectricity, and between anti- and ferromagnetic ordering was demonstrated recently by the simultaneous realization of ferromagnetism and ferroelectricity in strained EuTiO$_3$ thin films \cite{eutio3}. Magnetic quantum paraelectrics are therefore important for both fundamental physics and technological applications.

\begin{figure}
\includegraphics[clip,width=7.5cm]{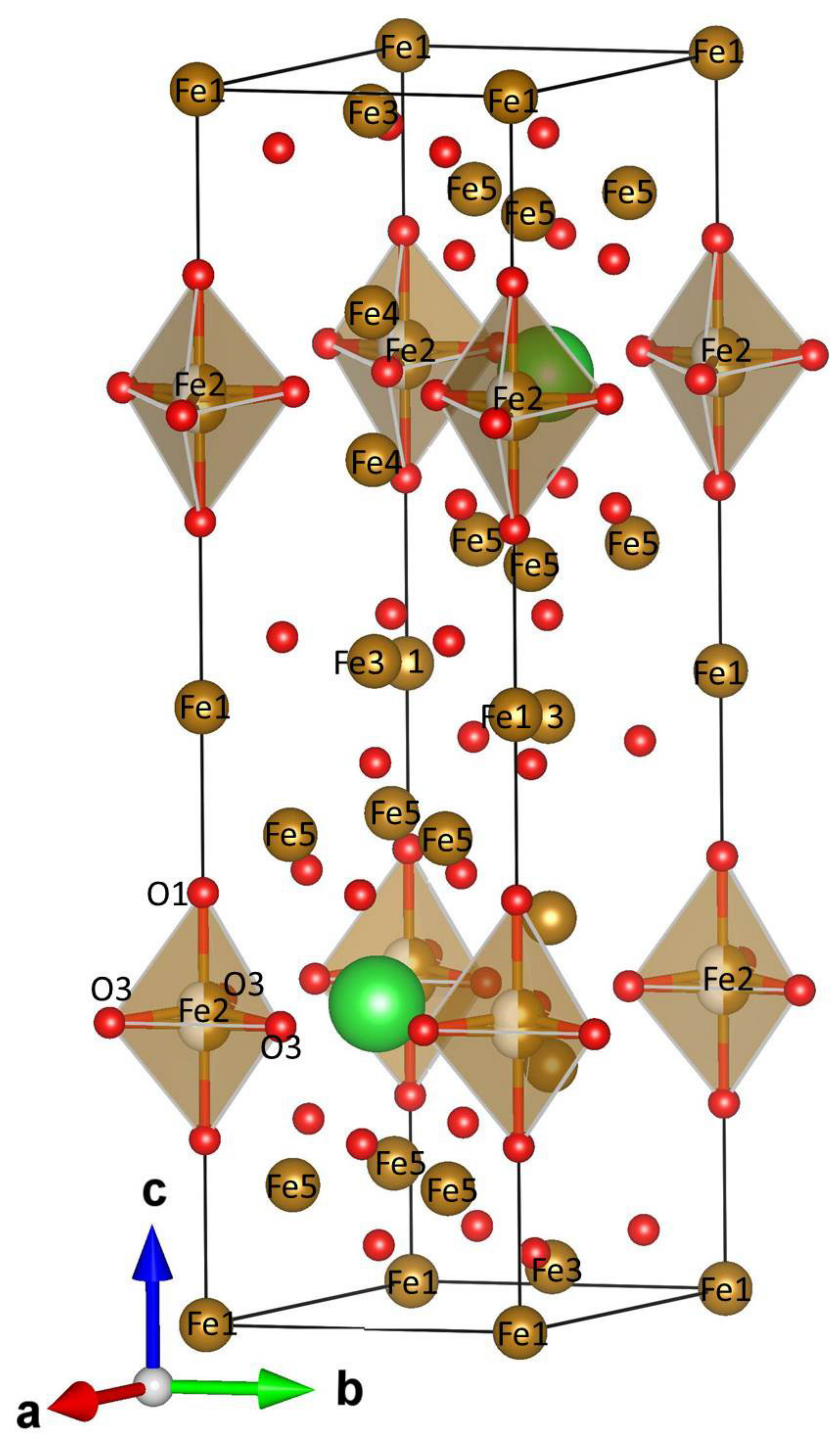}
\caption{(Color online) The nuclear structure of BaFe$_{12}$O$_{19}$ showing the five Fe sites. The Fe2 atoms occupy the trigonal bipyramidal sites. Fe$^{3+}$ ions at TBP sites displace away from the mirror plane. The brown balls represent Fe and are labeled by number. The red and green balls represent O and Ba, respectively. The partially filled brown balls represent half occupied sites.}
\label{nuc}
\end{figure}

Recently BaFe$_{12}$O$_{19}$, a well-known ferrite  permanent magnet with $T_C$ above 720 K, was proposed to exhibit quantum paraelectric behavior \cite{shen}. BaFe$_{12}$O$_{19}$  is an important member of the hexagonal ferrites, also known as hexaferrites, discovered in 1950s \cite{went, wijn, braun, jonker, smit}. It has been widely used as a permanent magnet, in magnetic recording media, in microwave devices, and also for fabricating multiferroic devices \cite{smit, koji, pullar, tokunaga, tan}. BaFe$_{12}$O$_{19}$  crystallizes in the centrosymmetric magnetoplumnite structure with space group $P6_3/mmc$. As shown in Fig. \ref{nuc}, the magnetic Fe$^{3+}$ ions occupy five different crystallographic sites: octahedral sites ($2a$, $4f$, and $12k$), tetrahedral ($4f$) sites, and trigonal bipyramidal (TBP) $4e$ sites. In the hexaferrites, all the structures can be obtained by stacking the S, R, and T blocks. Here the S, R, and T blocks are made of two, three, or four oxygen layers, respectively. The $\ast$ symbol means that the preceding block is turned 180$^\circ$ around the hexagonal $\emph{c}$ axis. BaFe$_{12}$O$_{19}$ has a stacking sequence of SRS$^\ast$R$^\ast$S.  A collinear ferrimagnetic structure forms below $T_C$ = 723 K with the moment direction parallel to the hexagonal $\emph{c}$ axis \cite{gorter,magtc,pwd}. The Fe$^{3+}$ magnetic moments at two $4f$ sites are antiparallel to those at the other sites ($2a$, $4e$, and $12k$), with a high spin state, which yields a net moment of up to 40 $\mu \rm_B$ per unit cell.

The TBP Fe$^{3+}$ ions show unusual off-mirror-plane (OMP) displacements which result in local electric dipoles \cite{PRX,shen,tan}. Neutron powder diffraction measurements suggested that the OMP displacement of TBP Fe$^{3+}$ ion is 0.26 {\AA} in the paramagnetic phase at 743~K, and the TBP Fe$^{3+}$ ions freeze in the mirror-plane at 4.2~K \cite{pwd}. The freezing of the TBP Fe$^{3+}$ ions at low temperatures agrees with a M{\"o}ssbauer study reported by Mamalui, \textit{et al.}\cite{moss1} However, in a later M{\"o}ssbauer study, the OMP displacement was found to be nonzero and to increase linearly upon warming in the temperature range from 4~K to 300~K. The room temperature OMP displacement of 0.185 {\AA} is consistent with an X-ray measurement \cite{xray} and another M{\"o}ssbauer measurement \cite{moss3}. It is still not clear whether the TBP Fe$^{3+}$ ions freeze in the mirror plane at low temperatures. Very recently, a Monte Carlo simulation predicted that antiferroelectric order takes place below 3~K in BaFe$_{12}$O$_{19}$ \cite{PRX}. However, a later dielectric constant measurement did not find electric dipole order above 2~K, but did reveal a plateau feature that was attributed to the suppression of the long range electric dipole order by quantum fluctuations \cite{shen}. Therefore, it is important to resolve the discrepancy about the OMP displacements of the TBP Fe ions at low temperatures in order to understand the low temperature dielectric properties.

In this paper, we report a detailed neutron diffraction study of BaFe$_{12}$O$_{19}$ single crystals grown by a high pressure floating zone technique. The neutron diffraction study shows that the OMP displacement of TBP Fe$^{3+}$ ions at 4~K is nonzero, and is about half of the displacement value at 300 K. The reduced OMP displacement is accompanied by a displacement of the apical oxygen ions toward the TBP Fe ions.  We also studied the dielectric properties of BaFe$_{12}$O$_{19}$ single crystals at temperatures down to 0.4~K in order to look for possible antiferroelectric order. The temperature dependence of dielectric constant shows a non-classical $T^2$ dependence of the inverse dielectric function from 10-20~K and an anomalous upturn below 1.40~K extending to the millikelvin range. Our results provide strong evidence that BaFe$_{12}$O$_{19}$ is simultaneously a ferrimagnet and a quantum paraelectric.

\begin{figure} \centering \includegraphics [width = 0.47\textwidth] {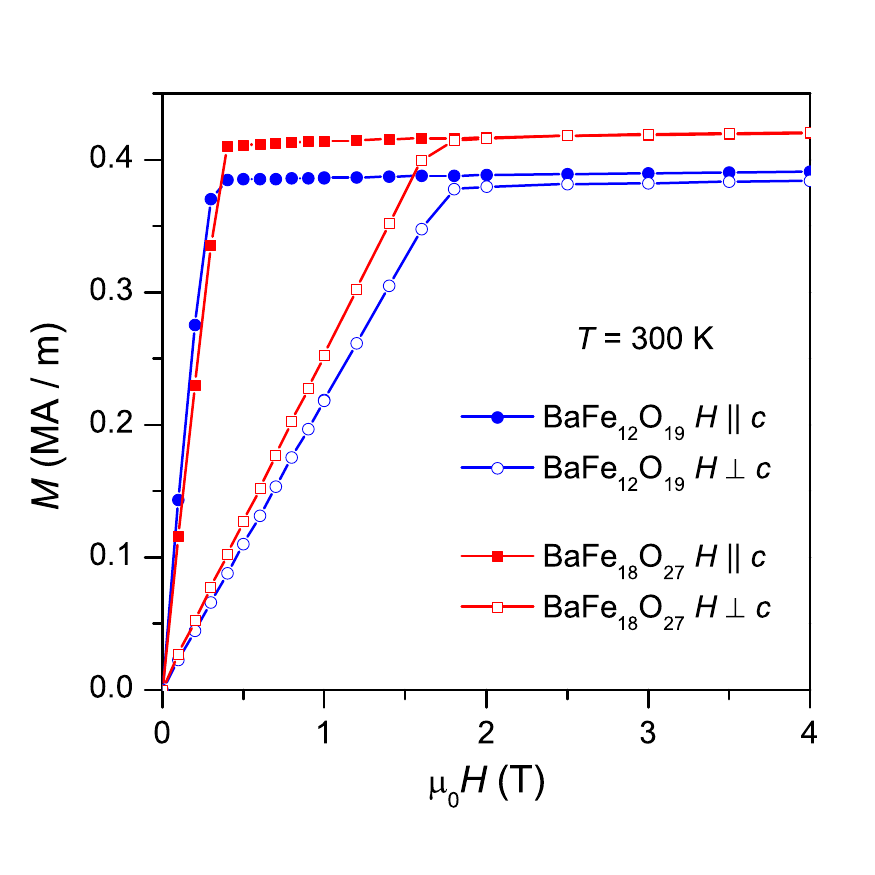}
\caption{(color online) Measured magnetization vs applied magnetic field for single crystals of BaFe$_{12}$O$_{19}$ and BaFe$_{18}$O$_{27}$ both parallel and perpendicular to the $\emph{c}$-axis, the easy axis of magnetization.}
\label{MH-1}
\end{figure}

\section{Experiments}

BaFe$_{12}$O$_{19}$ powder was made by conventional solid-state reaction of appropriate mixtures of BaCO$_3$ (Alfa,  99.99\%) and Fe$_2$O$_3$ (Alfa, 99.995\%) at 1100$^{\circ}$C for 72 hours with intermediate grindings. The BaFe$_{12}$O$_{19}$ feed rods used for the single crystal growth were made by pressing powder under hydrostatic pressure and then annealing these rods at 1200$^{\circ}$C in air.  Single crystals with dimensions of 5 mm in diameter by 60 mm long were grown by the floating zone method in flowing O$_2$ atmosphere at a pressure of 100~atm. The gas flow rate was 0.2~L/min. The high pressure floating zone growth was performed using a vertical optical image furnace (Model HKZ, SciDre, GmbH, Dresden) equipped with a 5~KW xenon arc lamp. During crystal growth, the upper and lower rods were counter-rotated at a relative speed of 25~rpm in order to maintain a homogeneous melt. The crystal was grown at a rate of 6~mm/h.

Measurements of the magnetic properties were performed in a Quantum Design (QD) MPMS. The temperature dependence of the electrical resistivity and heat capacity  were measured using a QD Physical Property Measurement System. A conventional four-probe technique was employed for the resistivity measurements. Pt wires were attached to rectangular bars cut from the bulk crystal using silver epoxy.

For the dielectric function measurements, a single-crystal sample was polished into a plate with typical dimensions of 4 $\times$ 4 $\times$ 0.1 mm$^3$ to have a parallel-plate capacitor geometry. The plate was cut perpendicular to the $c$-axis. The electrical contacts were made by coating the two flat surfaces with silver epoxy. The capacitance of the sample was measured with an Andeen-Hagerling AH-2700A capacitance bridge, and the dielectric function was obtained by approximating the sample as an ideal parallel-plate capacitor.

Single-crystal neutron diffraction measurements were performed on the HB-3A four-circle diffractometer at the High Flux Isotope Reactor, Oak Ridge National Laboratory. A neutron wavelength of 1.003 {\AA} was used, obtained from a bent perfect Si-331 monochromator \cite{hb3a}. Rietveld refinements were performed using FullProf  \cite{fullprof}. A maximum entropy method (MEM) analysis was carried out to obtain the nuclear density map using the Dysnomia software package\cite{Dysnomia}.

\section{Results}

\subsection{High pressure floating zone growth}

Under one atmosphere of oxygen, BaFe$_{12}$O$_{19}$ decomposes into BaFe$_{18}$O$_{27}$ (the so-called $W$-phase) and other oxides before melting. This incongruent melting  can be suppressed by high pressure oxygen, and  congruently melting behavior was reported above about 50 atm of oxygen \cite{growth1, growth2}.  Since the capability of maintaining 50 atm oxygen pressure is absent in most growth laboratories, BaFe$_{12}$O$_{19}$ crystals and doped variants have been mainly grown with flux techniques.

The Model HKZ high pressure image furnace enables crystal growth in controlled atmosphere at pressures up to 150 atm.  The high pressure capability can be used to suppress the evaporation loss of volatile components and tune the oxygen content in oxides. The floating zone growth of BaFe$_{12}$O$_{19}$ in high oxygen pressure provides a good example that high oxygen pressure can be used to modify the phase stability in $P-T-x$ space.

To obtain a high quality crystal, growth parameters such as oxygen pressure, flow rate, growth rate, and the rotation speed of both feed and seed rods, and temperature profile, are optimized. It was found that an oxygen pressure of 50~atm is necessary to obtain the right phase and maintain a stable growth. The effect of the oxygen pressure was tested and once the growth was performed in oxygen pressure below 50 atm, the as-grown ingot contains iron oxides (Fe$_2$O$_3$ and/or Fe$_3$O$_4$ depending on the oxygen pressure and temperature) and BaFe$_{18}$O$_{27}$ crystals with the largest dimension up to 8 mm.  Room temperature X-ray diffraction of pulverized single crystals confirmed that BaFe$_{18}$O$_{27}$ is phase pure with the correct lattice parameters.

A stable growth was obtained for BaFe$_{12}$O$_{19}$ in an oxygen pressure of 50-100~atm with a steady flow rate of 0.2~L/min. Crystals grown with a growth rate of 4-10~mm/h are of similar quality. The rotation speed of feed and seed rods shows little effect on the growth stability. The room temperature X-ray powder diffraction pattern was collected on pulverized single crystals. No extra reflections were observed. The crystal quality was further confirmed with X-ray back Laue diffraction.

Room temperature magnetization measurements were carried out on both BaFe$_{12}$O$_{19}$ and BaFe$_{18}$O$_{27}$ crystals. The results are shown in Fig. \ref{MH-1}. Measurements were conducted both with the field applied along the easy axis of magnetization, the $\emph{c}$-axis in these hexagonal materials, and with the field perpendicular to this axis, in the $ab$-plane. The measured saturation magnetization for the BaFe$_{12}$O$_{19}$ and BaFe$_{18}$O$_{27}$  crystals is 0.39 MA/m (73 emu/g, 14.5 $\mu_B$/F.U.) and 0.42 MA/m (80 emu/g, 22.4 $\mu_B$/F.U.), respectively. These values are in good agreement with accepted literature values of 0.38 MA/m for BaFe$_{12}$O$_{19}$ and 0.41 MA/m for BaFe$_{18}$O$_{27}$ \cite{skomski}. The anisotropy field $H_A$ can be estimated by noting the field at which the magnetization saturates in the hard direction. The data in Fig. \ref{MH-1} gives $\mu_0H_A$ = 1.75 T for BaFe$_{12}$O$_{19}$ and 1.66 T for BaFe$_{18}$O$_{27}$. These values are somewhat larger than the literature values of 1.7 and 1.5 T, respectively \cite{skomski}. This may be due to demagnetization effects, which were unaccounted for here \cite{skomski}.

\begin{figure}
\includegraphics[clip,width=7.5cm]{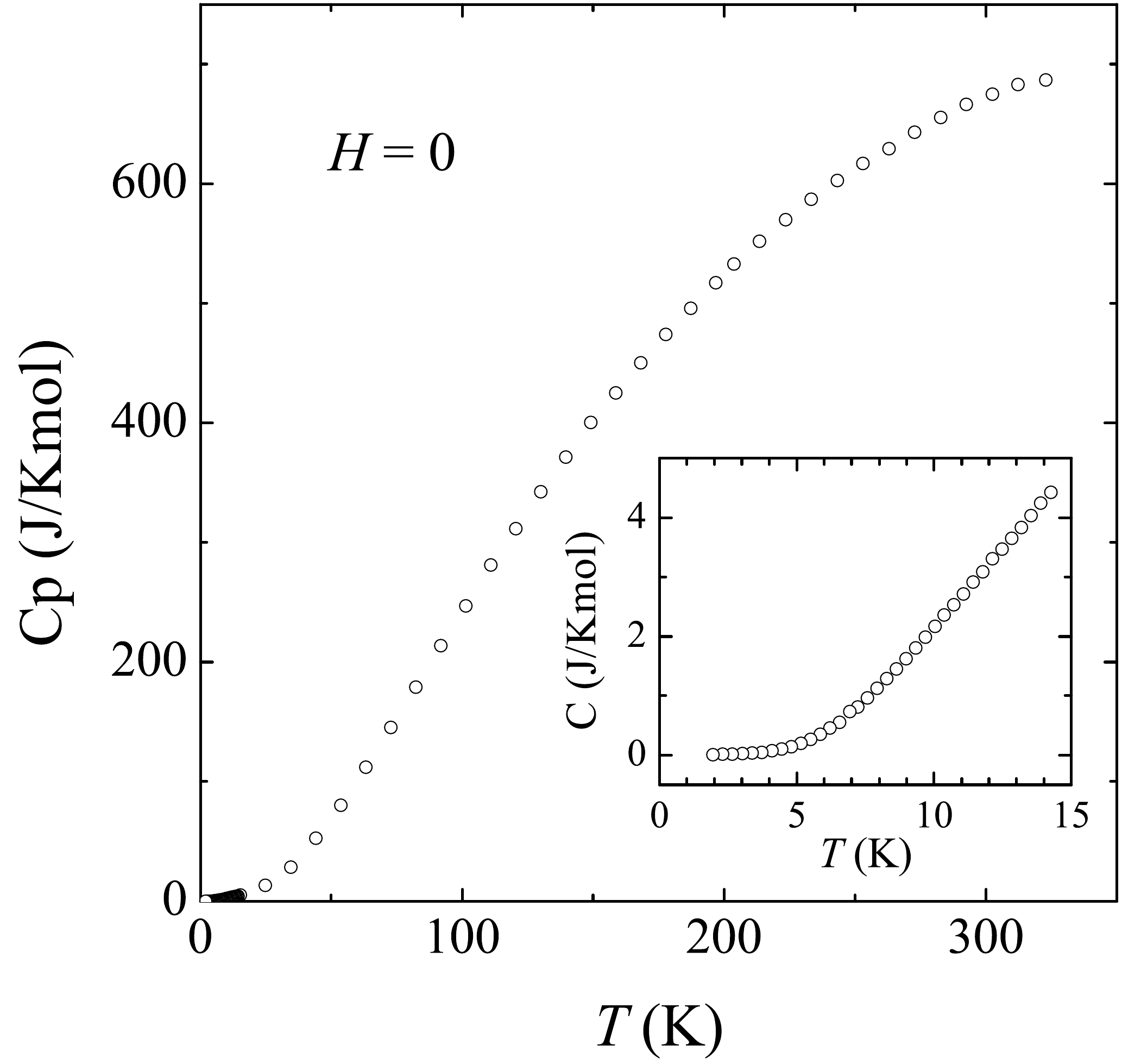}
\caption{(Color online) Temperature dependence of specific heat in zero field. The inset shows the region below 15 K.}
\label{cp}
\end{figure}

\begin{figure}
\includegraphics[clip,width=7.5cm]{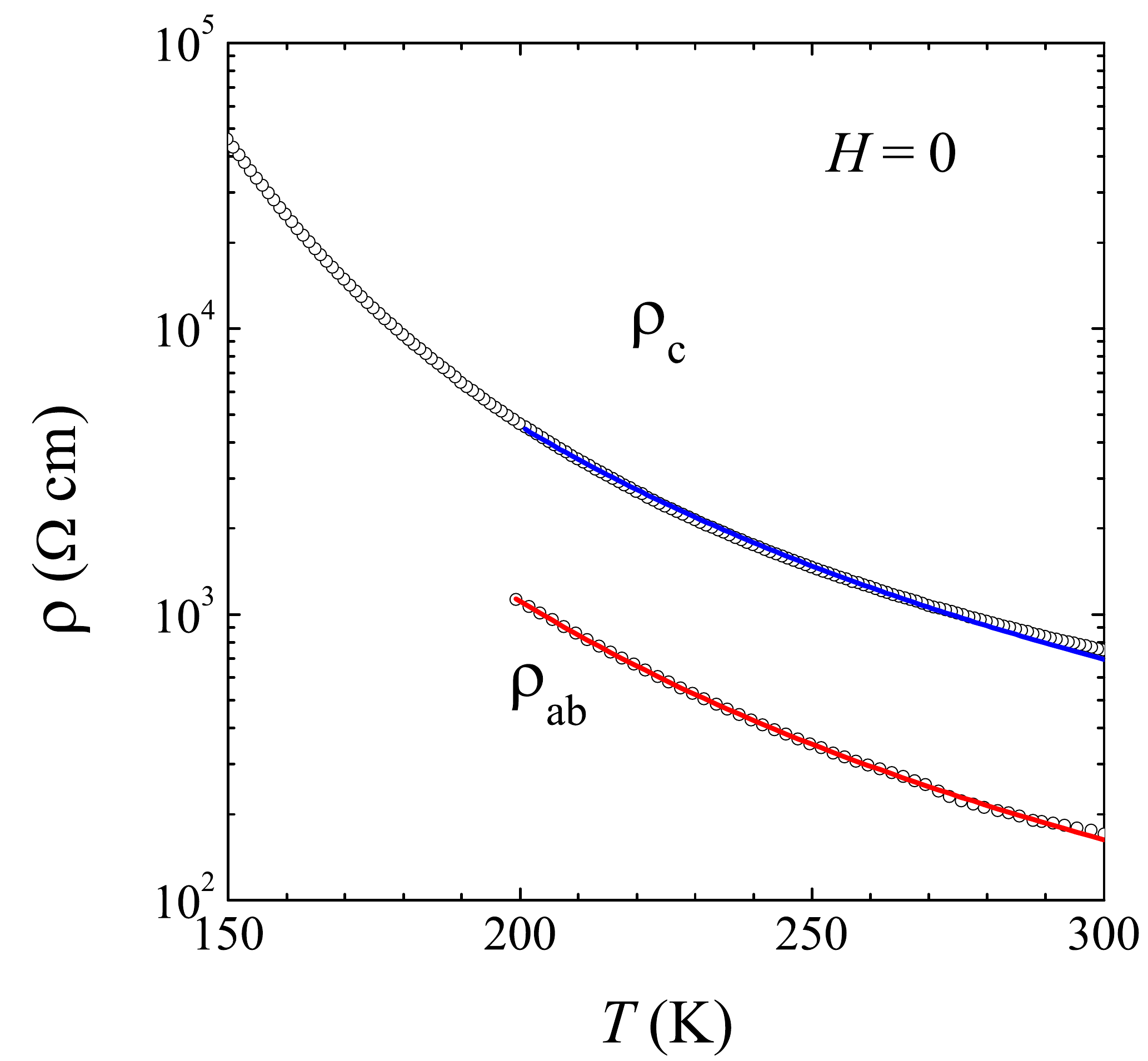}
\caption{(Color online) Temperature dependence of in-plane resistivity in zero field. The solid lines are fit using the activation energy model as described in the text.}
\label{rt}
\end{figure}

\subsection{Specific heat and anisotropic transport properties}

Figure \ref{cp} shows the temperature dependence of the specific heat of a BaFe$_{12}$O$_{19}$ single crystal measured in the temperature range 1.90~K-300~K. The temperature dependence and magnitude agree well with the data reported on flux-grown crystals \cite{shen}. As shown in the inset of Fig. \ref{cp}, there is no observable anomaly in our measurement down to 1.9~K, indicating the absence of long-range dipolar order. This result is in contrast to the most recent Monte Carlo simulation on BaFe$_{12}$O$_{19}$, which predicted antiferroelectric order at 3 K \cite{PRX}.  Measurement in an applied magnetic field of 8~T along $\emph{c}$ shows no field effect, which indicates that neither electric dipole order nor the magnetic order can be tuned by an applied magnetic field of 8~T or less along $\emph{c}$.

The temperature dependence of electrical resistivity was measured in the $ab$ plane and along the $c$ axis below room temperature as shown in Fig. \ref{rt}. Due to the large resistance of the crystal, the measurement was only feasible above 150 K (200~K) in the case of $\rho \rm_c$ ($\rho \rm_{ab}$). It can be seen that BaFe$_{12}$O$_{19}$ is insulating with strong anisotropy. The in-plane electrical resistivity $\rho \rm_{ab}$ is one order smaller than the $c$-axis electrical resistivity $\rho \rm_c$. The resistivity can be fitted well using activation energy model $\rho = \rho_0\exp({\Delta/T})$ between 200-300 K. The fitted parameters are $\rho_0$ = 3.5 $\Omega$ cm and $\Delta$ = 1153 K for $\rho \rm_{ab}$ and $\rho_0$ = 16 $\Omega$ cm and $\Delta$ = 1127 K for $\rho \rm_c$.

\subsection{Neutron single crystal diffraction}
A high quality BaFe$_{12}$O$_{19}$ single crystal was selected for neutron single crystal diffraction. More than 850 reflections were collected at 4~K and 295~K, respectively. Rietveld refinements were performed to determine the nuclear and magnetic structures. The refined atomic positions and magnetic moments are listed in Table \ref{strs}. Selected bond distances and angles are presented in Table \ref{bond} and \ref{angle}. Upon cooling from 295~K to 4~K two main changes were observed in the vicinity of the TBP sites: (1) the apical O1 shifts along $\emph{c}$ towards the Fe2 ion by 0.021(3)~{\AA} (see Fig. \ref{nuc}). The distance between the O1 above and the O1 atom below Fe2 is reduced by 0.042(6)~\AA; and (2) the TBP Fe2 displaces away from the mirror plane at $z$=0.25 by 0.177(4)~{\AA} and 0.098(9)~{\AA} at 295~K and 4~K, respectively. The OMP displacement of the TBP Fe2 is reduced upon cooling from 295~K to 4~K by 0.079(7)~\AA, much larger than the O1 shift, 0.021(3)~\AA, which is consistent with the notion that the TBP electric dipoles arise from off-centering into double wells.

\begin{figure}
\includegraphics[clip,width=7.5cm]{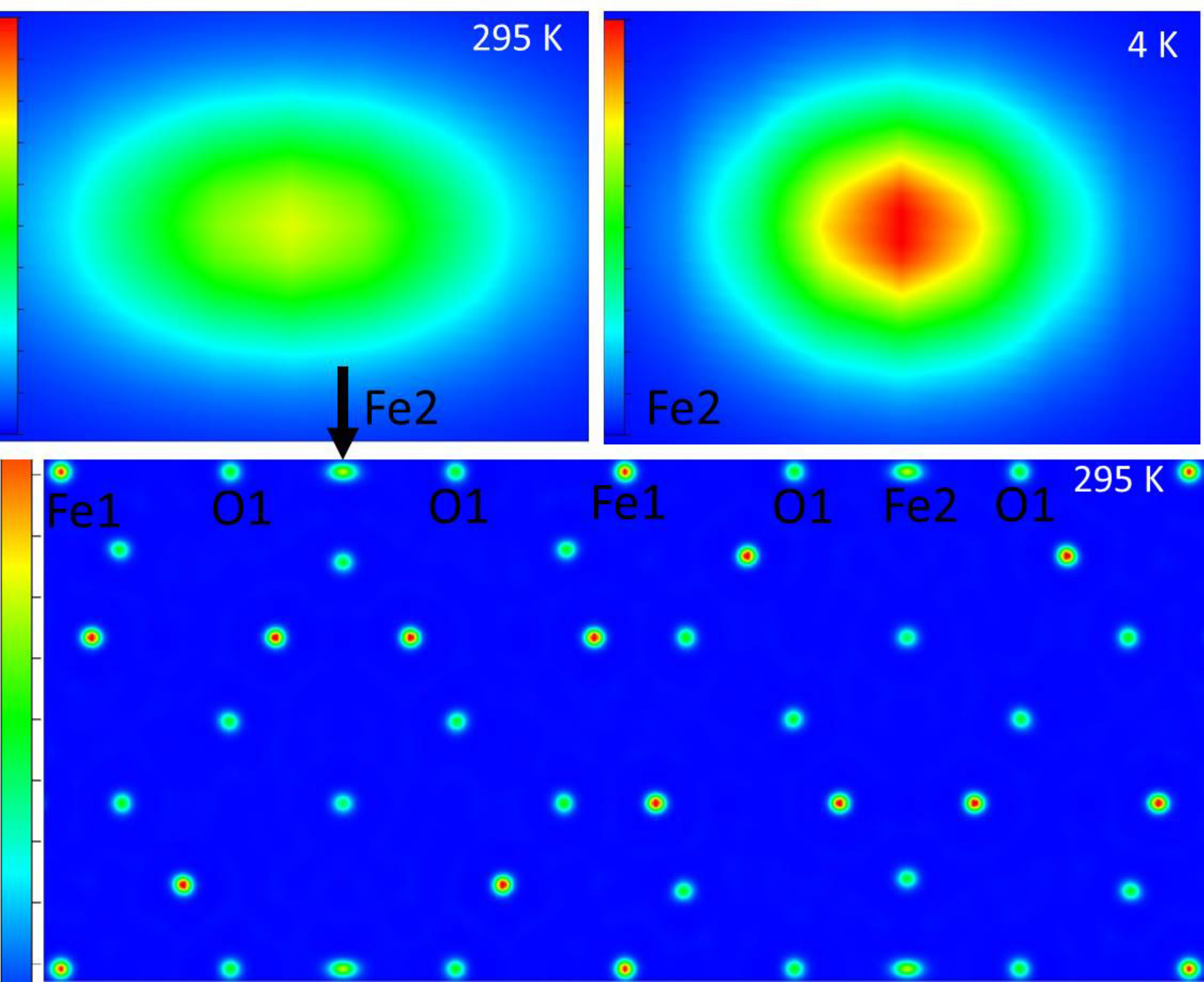}
\caption{(Color online) Nuclear density maps of BaFe$_{12}$O$_{19}$: the bottom shows the full unit cell in the crystal plane (1 1 0) at the origin. The most intense sites are for Fe$^{3+}$ ions with the largest neutron scattering length. The top two panels are the nuclear density maps around the TBP Fe sites at 295~K (left) and 4~K (right).}
\label{dens}
\end{figure}

To further understand the OMP displacements,  we performed a maximum entropy method (MEM) analysis. As shown in the nuclear density map obtained by the MEM analysis (Fig. \ref{dens}), only TBP Fe ions present a large displacement along the crystallographic $\emph{c}$ axis, and this displacement is reduced when cooling from 295~K to 4~K. The displacements of other atoms including the oxygen ions (O1 and O3) enclosing the TBP Fe are isotropic, exhibiting conventional thermal vibrational features. Therefore, a split TBP site along the $\emph{c}$ axis is necessary for both temperatures. The nuclear density at TBP sites does not show two maxima at either temperature because our neutron diffraction data were collected under the  $\emph{q}$ limit of 8~\AA$^{-1}$. Collection of higher $\emph{q}$ reflections will be necessary to resolve the expected maxima.

The observed OMP displacement at 295 K in our study, 0.177(4) \AA, agrees well with the previously reported values in a range of 0.16-0.185 {\AA} obtained by X-ray single crystal diffraction and M{\"o}ssbauer measurements \cite{xray, moss2, moss3}. At 4 K, we observed a reduced OMP displacement of 0.098(9) \AA, which is consistent with the latest M{\"o}ssbauer study \cite{moss2}; i.e., there is no freezing of the TBP Fe$^{3+}$ ions at the mirror plane as reported in the Ref. \cite{pwd, moss1}. At the TBP sites, there are two bond lengths for Fe2-O1 along $\emph{c}$ axis due to the OMP displacement of Fe2: 2.486(5)~{\AA} and 2.134(5)~{\AA} at 295~K, 2.386(9)~{\AA} and 2.192(9)~{\AA} at 4 K (see Table \ref{bond}), respectively. Both bonds are longer than the rest Fe-O bonds at each temperature. Obviously, these two Fe2-O1 bonds show a significant temperature dependence. In contrast, the Fe2-O3 bonds in the $\emph{ab}$-plane are much shorter with a length of 1.86~{\AA} and show no temperature dependence within the error bar.

The bond lengths of in-plane Fe2-O3 bonds are close to those of the Fe-O bonds (Fe3-O4) in the tetrahedral sites. In fact, each bipyramid can be viewed as two face-sharing tetrahedra, where the TBP Fe$^{3+}$ hops from one tetrahedron to the other, forming an electric dipole. In order to look for possible long range order of the dipoles, high resolution diffraction data were collected with both X-ray and neutron single crystal diffraction. No  peak breaking the reflection condition $l$=2n within the space group $P 6_3/m m c$ was observed with neutron single crystal diffraction at 4 K. The absence of superlattice peaks was further confirmed with X-ray single crystal diffraction measurements performed in the temperature range 28 K-300 K (results not shown). Thus, no long range antiferroelectric order forms above 4~K in BaFe$_{12}$O$_{19}$, consistent with the conclusion from the specific heat measurements.

\begin{widetext}

\begin{table}

\caption{Refined structures of BaFe$_{12}$O$_{19}$ from the single crystal neutron diffraction are presented in space group $P 63/m m c$. The data was collected  with a wavelength of 1.003~{\AA} at HB-3A at HFIR. The lattice parameters are $\emph{a}$ = 5.8948(8)~{\AA} and $\emph{c}$ = 23.202(3)~{\AA} at 295 K, $\emph{a}$ =  5.8966(8)~{\AA} and $\emph{c}$ = 23.193(3)~{\AA} at 4~K.}
\begin{tabular}{c|c|ccccc|cccccccc}
\hline
~~~~~~ & && ~~~~$T$ = 295~K & & &&& &~$T$ = 4~K ~&&& \\
\hline
atom & $type$ & $x$ & $y$ & $z$ & $B_{iso}$ & $M_z$($\mu_B$) & $x$ & $y$ & $z$ & $B_{iso}$ & $M_z$($\mu_B$)\\
\hline
Ba1 & Ba & 0.6667(0) & 0.3333( 0) &  0.25000( 0) & 0.32(8)&& 0.6667(0) & 0.3333( 0)&  0.25000( 0)&  0.08(8) & &  \\
Fe1 & Fe & 0.0000(0) & 0.0000( 0) &  0.00000( 0) & 0.24(4)& 4.36(26)& 0.0000(0) & 0.0000( 0)&  0.00000( 0)&  0.10(4) & 4.67(22) \\
Fe2 & Fe & 0.0000(0) & 0.0000( 0) &  0.25764(19) & 0.30(7)& 3.54(26)& 0.0000(0) & 0.0000( 0)&  0.25422(37)&  0.08(6) & 3.92(21) \\
Fe3 & Fe & 0.3333(0) & 0.6667( 0) &  0.02724( 9) & 0.25(3)&-4.22(20)& 0.3333(0) & 0.6667( 0)&  0.02726( 9)&  0.11(3) &-4.36(19) \\
Fe4 & Fe & 0.3333(0) & 0.6667( 0) &  0.19031( 9) & 0.33(3)&-4.05(14)& 0.3333(0) & 0.6667( 0)&  0.19042( 9)&  0.15(3) &-4.33(13) \\
Fe5 & Fe & 0.1688(3) & 0.3375( 5) &  0.89171( 4) & 0.29(2)& 3.34( 9)& 0.1687(3) & 0.3373( 5)&  0.89160( 4)&  0.12(2) &4.22( 8) \\
O1  & O  & 0.0000(0) & 0.0000( 0) &  0.15033(14) & 0.42(5)&& 0.0000(0) & 0.0000( 0)&  0.15125(15)&  0.30(5) &    & \\
O2  & O  & 0.3333(0) & 0.6667( 0) &  0.94529(16) & 0.38(5)&& 0.3333(0) & 0.6667( 0)&  0.94505(15)&  0.22(5) &    & \\
O3  & O  & 0.1818(7) & 0.3635(13) &  0.25000( 0) & 0.46(4)&& 0.1819(6) & 0.3637(13)&  0.25000( 0)&  0.29(4) &    &  \\
O4  & O  & 0.1564(5) & 0.3128( 9) &  0.05204( 8) & 0.38(3)&& 0.1564(4) & 0.3128( 9)&  0.05214( 8)&  0.24(3) &    & \\
O5  & O  & 0.5025(7) & 0.0049(13) &  0.14920( 7) & 0.39(3)&& 0.5023(6) & 0.0045(13)&  0.14939( 7)&  0.27(3) &    & \\
\hline
$Rf$&&&0.023 &&&&&0.035\\
$\chi^2$&&&0.20&&&&&0.26\\
\hline
\end{tabular}
\label{strs}
\end{table}
\end{widetext}

\begin{widetext}

\begin{table}
\caption{The bond lengths for each FeO$n$ polyhedra at 295~K and 4~K. The units are \AA.}
\begin{tabular}{c|c|c|c|c|c|c|c|c|c|c|c|c|c|c|c}
\hline
Octa&&&Bipyramid&&&Tetra&&&Octa&&&Octa& \\
\hline

bond   & 295~K &  4~K & bond & 295~K &  4~K &bond & 295~K &  4~K &bond & 295~K &  4~K &bond & 295~K &  4~K  &\\
\hline
Fe1$-$O4  &     2.001   &    2.001  &    Fe2$-$O1     &      2.486   &    2.386  &    Fe3$-$O2    &       1.901    &   1.906 & Fe4$-$O3        &   2.076  &     2.072   &   Fe5$-$O1     &      1.979   &    1.987 &\\
\hline
Fe1$-$O4  &     2.001   &    2.001  &    Fe2$-$O1     &      2.134   &    2.192  &    Fe3$-$O4    &       1.894    &   1.895 &  Fe4$-$O3        &   2.076  &     2.072   &   Fe5$-$O2     &      2.087   &    2.087 &\\
\hline
Fe1$-$O4  &     2.001   &    2.001  &    Fe2$-$O3     &      1.862   &    1.858  &    Fe3$-$O4    &       1.894    &   1.895 &  Fe4$-$O3        &   2.075  &     2.071   &   Fe5$-$O4     &      2.111   &    2.111 &\\
\hline
Fe1$-$O4  &     2.001   &    2.001  &    Fe2$-$O3     &      1.862   &    1.858  &    Fe3$-$O4    &       1.894    &   1.894 &  Fe4$-$O5        &   1.969  &     1.968   &   Fe5$-$O4     &      2.112   &    2.112 &\\
\hline
Fe1$-$O4  &     2.001   &    2.001  &    Fe2$-$O3     &      1.862   &    1.858  &              &                &         &  Fe4$-$O5        &   1.969  &     1.968   &   Fe5$-$O5     &      1.926   &    1.928  &\\
\hline
Fe1$-$O4  &     2.001   &    2.001  &               &              &           &              &                &         &  Fe4$-$O5        &   1.97   &     1.968   &   Fe5$-$O5     &      1.926   &    1.928 &\\

\hline
\end{tabular}
\label{bond}
\end{table}
\end{widetext}

\begin{table}
\caption{The Fe-O-Fe bond angles at 295~K and 4~K. The units are degrees.}
\begin{tabular}{c|c|c}
\hline
 bond angle& T=295~K      &           T=4~K       \\
\hline
 Fe1 - O4 -  Fe3 & 125.19(11)  &     125.11(9)      \\
 Fe1 - O4 -  Fe5 & 95.45(13)   &     95.50(11)      \\
 Fe1 - O4 -  Fe5 & 95.48(12)   &     95.51(11)      \\
 Fe2 - O1 -  Fe5 & 119.5(2)    &     120.0(5)       \\
 Fe2 - O1 -  Fe5 & 119.5(2)    &     120.0(5)       \\
 Fe2 - O1 -  Fe5 & 119.5(2)    &     120.0(5)       \\
 Fe2 - O1 -  Fe5 & 119.5(3)    &     120.0(5)       \\
 Fe2 - O1 -  Fe5 & 119.5(3)    &     120.0(5)       \\
 Fe2 - O1 -  Fe5 & 119.5(3)    &     120.0(5)       \\
 Fe2 - O3 -  Fe4 & 132.72(17)  &     135.2(2)       \\
 Fe2 - O3 -  Fe4 & 132.72(17)  &     135.2(2)       \\
 Fe2 - O3 -  Fe4 & 143.58(18)  &     141.2(2)       \\
 Fe2 - O3 -  Fe4 & 143.58(18)  &     141.2(2)       \\
 Fe3 - O2 -  Fe5 & 126.49(19)  &     126.41(19)     \\
 Fe3 - O2 -  Fe5 & 126.51(18)  &     126.43(18)     \\
 Fe3 - O2 -  Fe5 & 126.51(18)  &     126.43(18)     \\
 Fe3 - O4 -  Fe5 & 121.23(16)  &     121.27(15)     \\
 Fe3 - O4 -  Fe5 & 121.26(14)  &     121.28(12)     \\
 Fe4 - O5 -  Fe5 & 128.18(17)  &     128.20(19)     \\
 Fe4 - O5 -  Fe5 & 128.2(2)    &     128.22(15)     \\
\hline
\end{tabular}
\label{angle}
\end{table}

\begin{figure}
\includegraphics[clip,width=7.5cm]{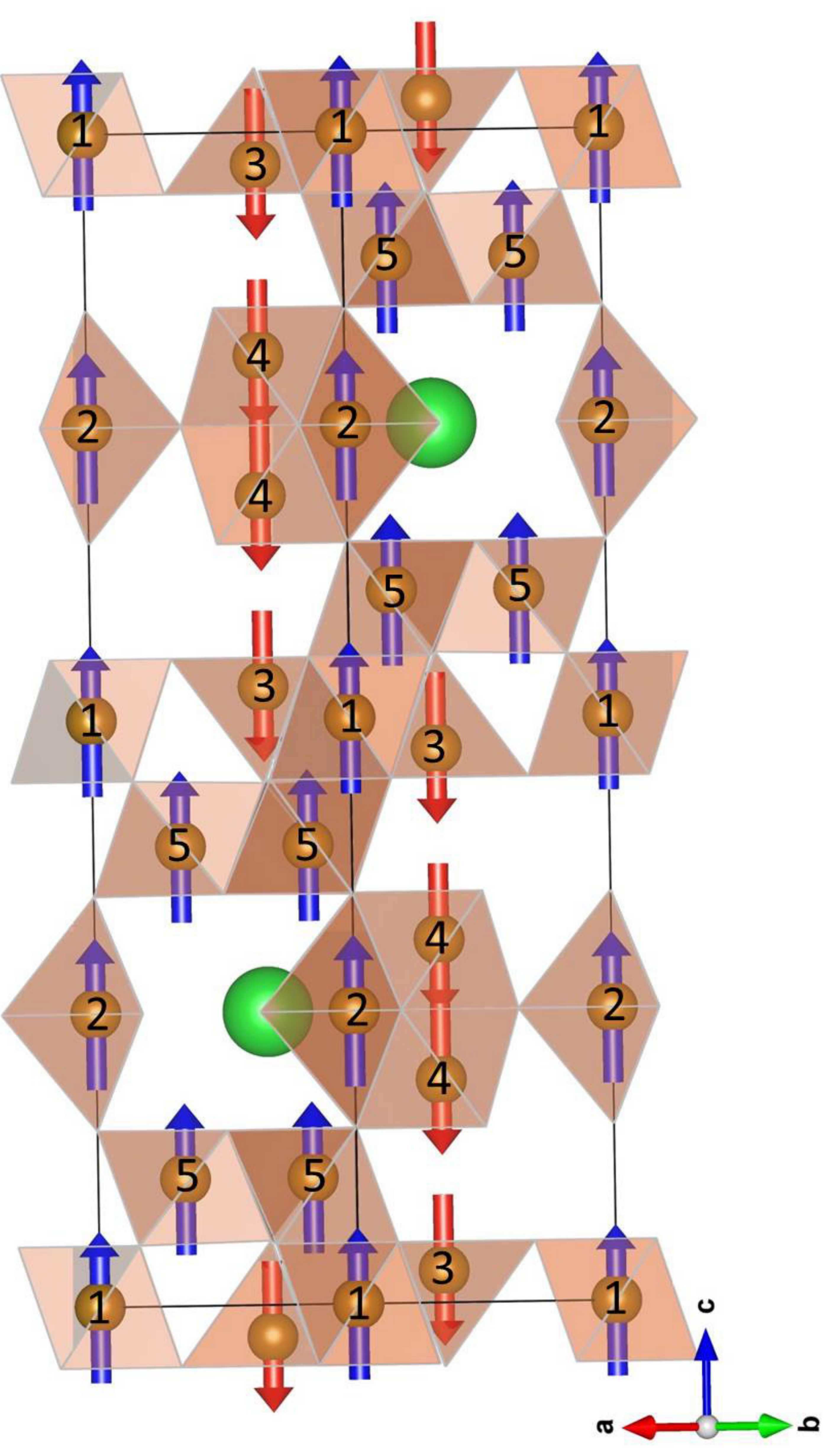}
\caption{(Color online) The magnetic structure of BaFe$_{12}$O$_{19}$. Fe atoms are labeled with numbers. The moment sizes are listed in Table I.}
\label{magstr}
\end{figure}

Since the neutron single crystal diffraction data were collected in the magnetically ordered state at temperatures far below $T_C$, we also solved the magnetic structure with all the spins along $\emph{c}$ axis (see Fig. \ref{magstr}). The overall magnetic structure is consistent with previous reports \cite{gorter, pwd}. Compared to the previous neutron powder diffraction study, our single crystal study reveals a smaller magnetic moment at each Fe$^{3+}$ site at 4~K (see Table \ref{strs}). The average moment size per Fe$^{3+}$ is 4.27(14) $\mu \rm_B$, smaller than the reported 4.6(2) $\mu \rm_B$. The net magnetic moment per Fe$^{3+}$ is 1.38(14) $\mu \rm_B$, which is also smaller than the reported 1.7(2) $\mu \rm_B$. Note, if all the Fe$^{3+}$ magnetic moments were constrained to have the same size in the refinements, a magnetic moment of 3.74(4) $\mu_B$ at 295~K and 4.28(4) $\mu_B$ at 4~K  are obtained, which gives a net magnetic moment of 14.96(48) $\mu_B$/F.U. at 295~K (consistent with the magnetization measurement) and 17.12(17) $\mu_B$/F.U. at 4~K.

As listed in Table I, the magnetic moment on all Fe sites increases when cooling from room temperature to 4~K. The increase is about 26\% for Fe5, much larger than the 3-10\% found for the other Fe sites. This increase might be related to the displacement of O1 toward Fe2. As shown in Fig. \ref{magstr}, the TBP shares O1 with the FeO$_6$ octahedron at Fe5 sites. The displacement of O1 toward Fe2 has little effect on the positions of other five oxygen in the Fe5O$_6$ octahedron but increases the bond length of Fe5-O1. At 295~K,  the Fe2 and Fe5 sites that bond with O1 show a comparable magnetic moment of 3.54(26) $\mu \rm_B$ and 3.34(9)$\mu \rm_B$, respectively, smaller than that on the other Fe sites. Upon cooling to 4~K, the magnetic moment on each Fe site becomes comparable to one another. This suggests that the local structural distortion affects both the local dipoles and the magnetic moments.

\begin{figure}
\includegraphics[clip,width=7.5cm]{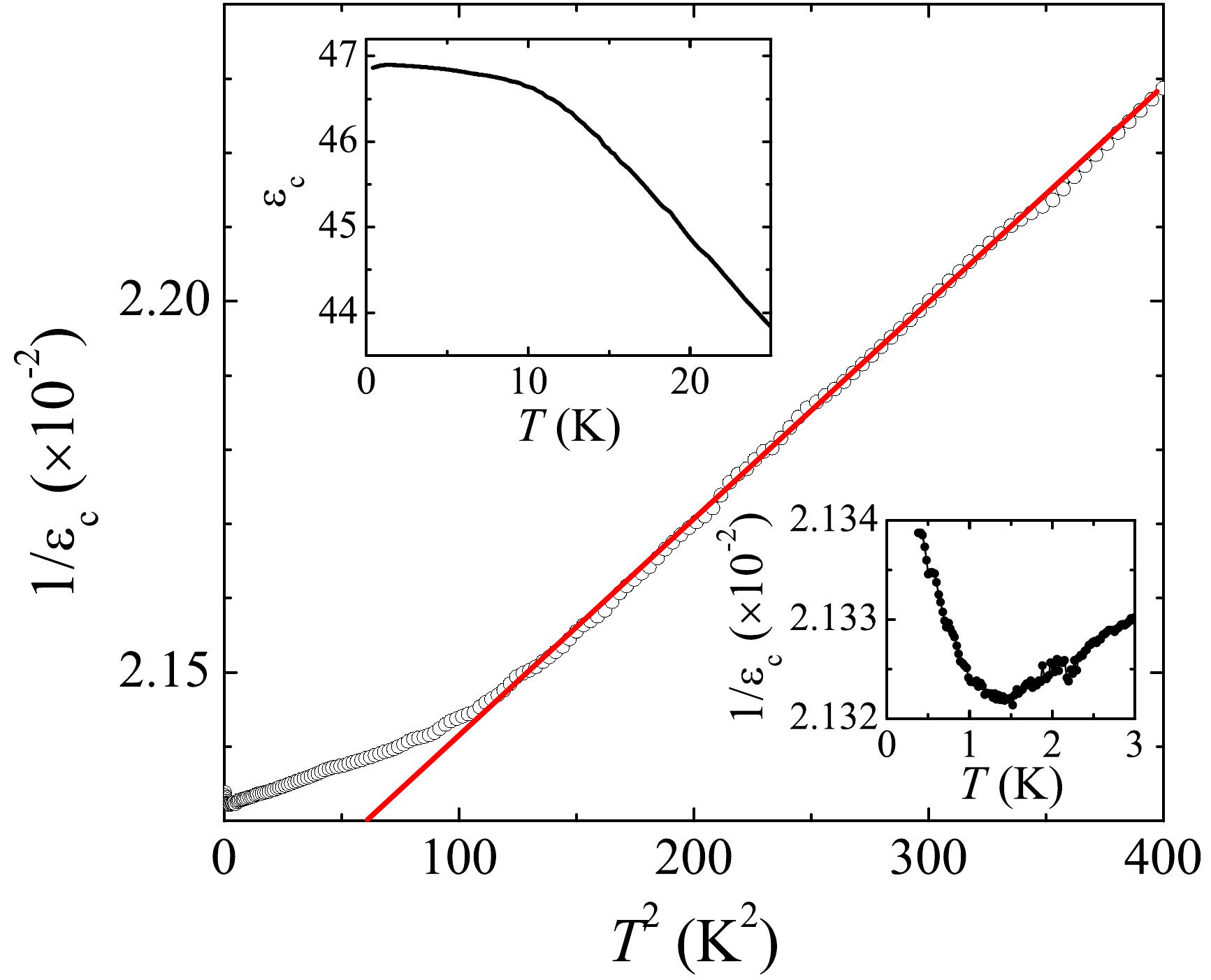}
\caption{(Color online) $T^2$ dependence of inverse dielectric function 1/$\varepsilon$ from 10-20~K. The upper inset shows the temperature dependence of $\varepsilon$. the lower inset highlights the anomalous upturn of 1/$\varepsilon$ below 1.4 K.}
\label{dielec}
\end{figure}

\subsection{Dielectric constant}

A recent first principles calculation predicted antiferroelectric (AFE) order at 3 K. However, no evidence was found for AFE order in the temperature dependence of the specific heat or permittivity measured on flux-grown crystals \cite{shen}. We therefore measured the dielectric function of our crystals grown by high pressure floating zone technique in the temperature range 0.4-25~K. Above 2~K the magnitude and temperature dependence of the permittivity agree well with the previous report \cite{shen}. The permittivity increases with decreasing temperature and saturates below about 10~K. No anomaly associated with the predicted AFE order was observed. Figure \ref{dielec} shows the inverse dielectric constant (1/$\varepsilon$) plotted against $T^2$. Three features are noteworthy: (1) 1/$\varepsilon$ follows a $T^2$ temperature dependence in the range 10-20~K; (2) below 10~K, 1/$\varepsilon$ deviates from this $T^2$ behavior, while $\varepsilon$ shows a plateau (see the top inset of Fig. \ref{dielec}); (3) below 1.4 K, 1/$\varepsilon$ shows an anomalous upturn (see lower inset of Fig. \ref{dielec}) which extends down to 0.4 K, the lowest temperature measured.

A regular Curie-Weiss law cannot describe the temperature dependence of the dielectric function, especially the plateau-like feature at low temperatures. In Ref. \cite{shen}, the mean-field Barrett formula was found to describe the temperature dependence of the $c$-axis dielectric permittivity well over the whole temperature range \cite{barrett}. Since the Barrett formula is often used to describe the dielectric properties of quantum paraelectrics, BaFe$_{12}$O$_{19}$ was, thus, proposed to be a magnetic quantum paraelectric compound \cite{shen}. Our dielectric constant data show similar temperature dependence as highlighted in the upper inset of Fig. \ref{dielec}, which indicates that the BaFe$_{12}$O$_{19}$ single crystals grown using high pressure floating zone technique used in this study show similar dielectric properties to flux-grown crystals.

The observed non-classical $T^2$ temperature dependence of the inverse dielectric function from 10-20 K, followed by the anomalous upturn below 1.4~K, are of particular interest. A similar behavior has been observed in SrTiO$_3$ and KTaO$_3$ quantum paraelectrics on the border of ferroelectricity \cite{quantum_paraelectric}. Both features can be understood in terms of the theory of quantum phase transitions when extended to include the effects of long-range dipolar interactions and the coupling of the electric polarization field with acoustic phonons. A similar temperature dependence of the dielectric properties of BaFe$_{12}$O$_{19}$ strongly suggests that BaFe$_{12}$O$_{19}$ is a critical quantum paraelectric system on the border of ferroelectricity.

\section{Discussions}

\begin{figure}
\includegraphics[clip,width=7.5cm]{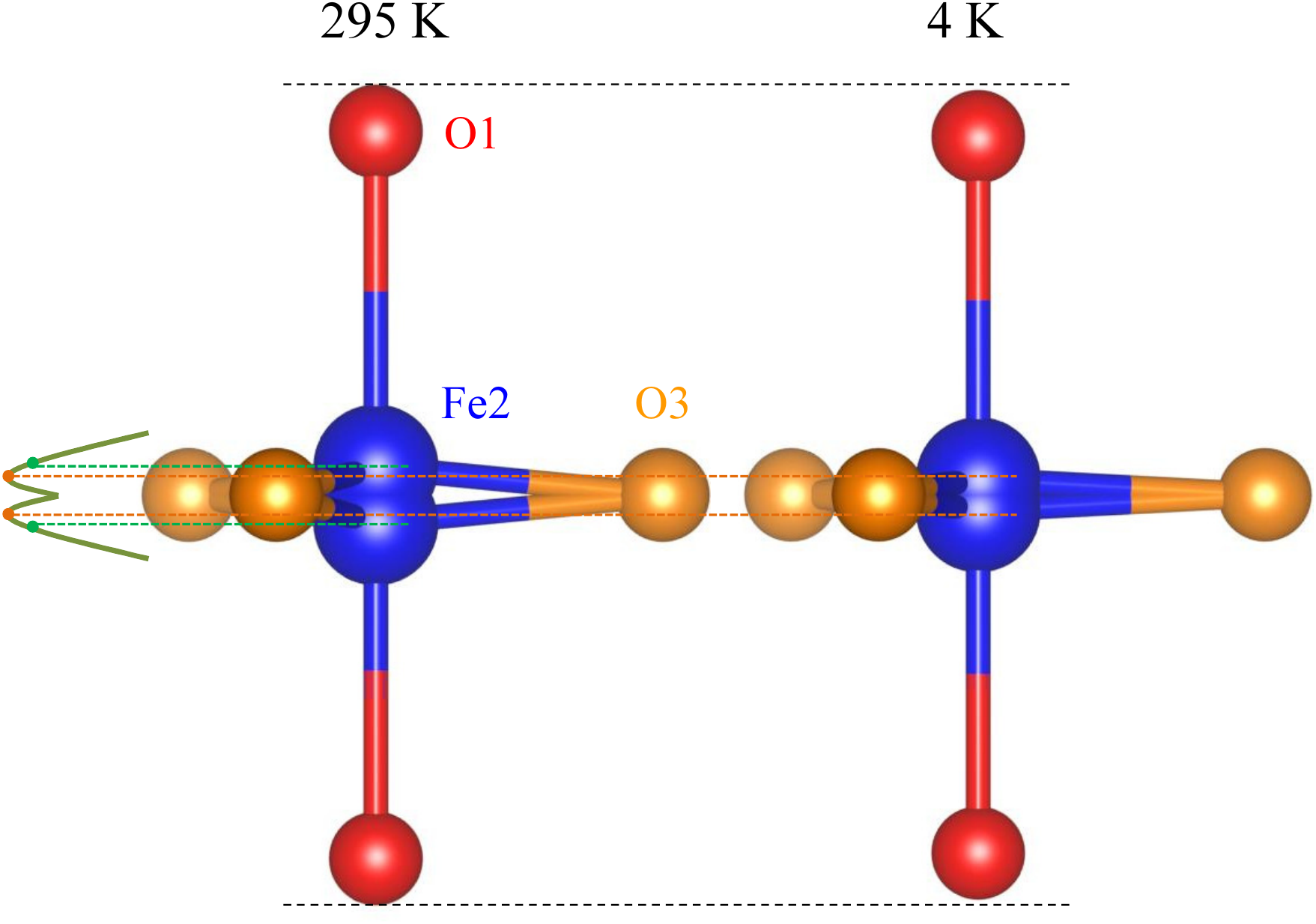}
\caption{(Color online) Comparison of local FeO$_5$ bipyramid structure at 295~K and 4~K. The black dashed lines represent the O1 position at 295 K. Green and orange dashed lines represent the center position of the Fe2 atom at 295~K and 4~K, respectively.}
\label{localstr}
\end{figure}

\begin{figure}
\includegraphics[clip,width=7.5cm]{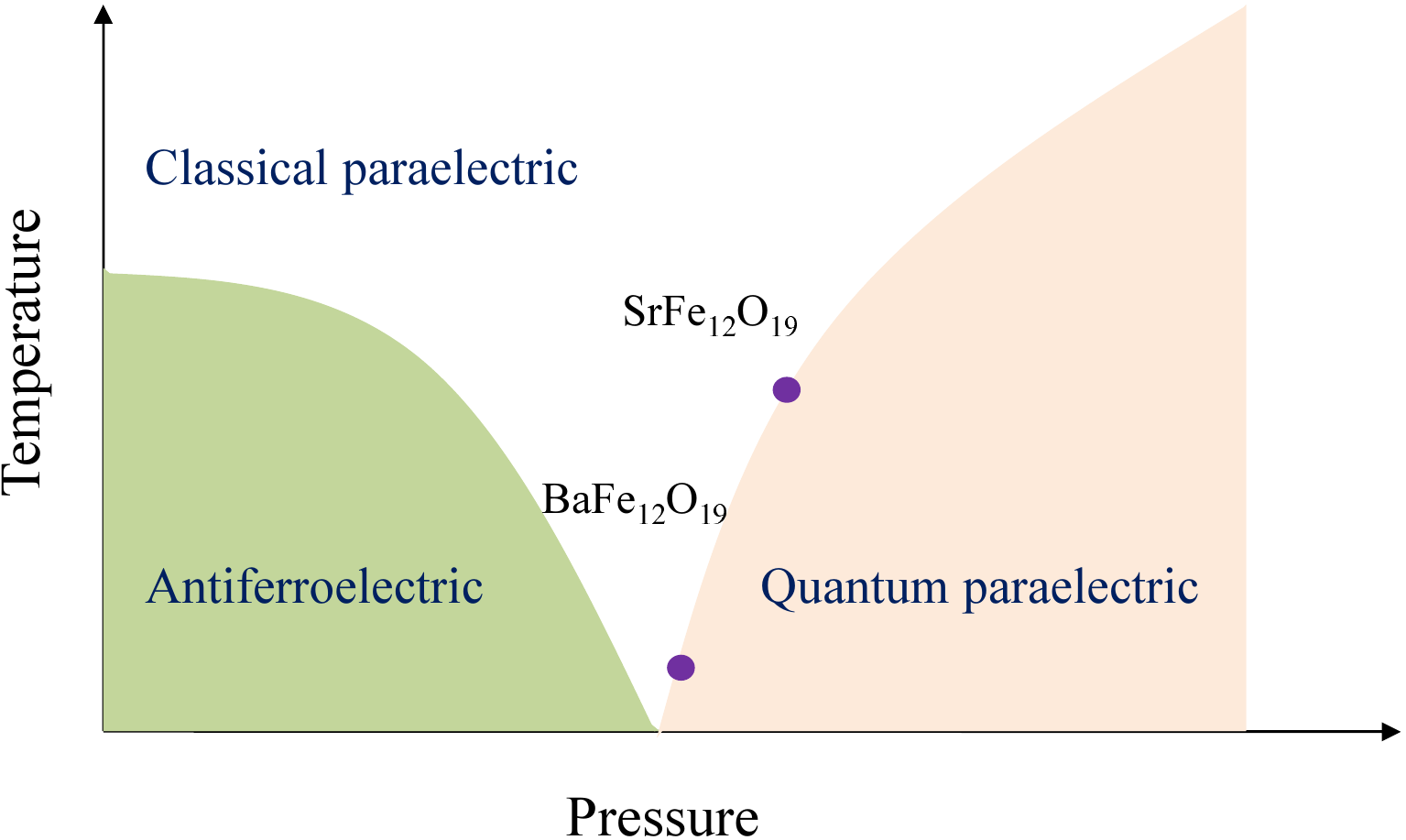}
\caption{(Color online) Schematic temperature-pressure phase diagram in $M$-type hexaferrites. The purple points are obtained by tracking the ending temperature of the $T^2$ dependence.}
\label{dgm}
\end{figure}

\subsection{High pressure floating growth}

The successful growth of BaFe$_{12}$O$_{19}$ single crystals demonstrates that floating zone growth under high pressure oxygen can extend the region of $P-T-x$ in which large, high quality crystals of certain materials can be grown. Floating zone is a crucible-free growth technique which avoids crystal contamination by reaction between the melt and crucible materials. Also, floating-zone crystals are free of flux inclusions since normally no flux is used. The floating zone growth of incongruently melting materials is rather challenging. It is normally performed with a so-called traveling solvent floating zone (TSFZ) technique, in which the melt has a different composition than the desired compound. With the help of the solvent, crystals can be grown at temperatures below the decomposition temperature. Crystal growth takes place while the solvent dissolves the feed materials and precipitates on top of the seed rods. The dissolving and precipitating from the solvent are usually slow, thus crystal growth with TSFZ technique is generally performed with a very slow growth rate, typically less than 1~mm/h. On the other hand, the solvent used normally has a composition different from the feed rod, thus the as-grown crystals can be contaminated by the solvent as flux inclusions. For the growth of BaFe$_{12}$O$_{19}$, the applied high oxygen pressure prevents decomposition at high temperatures. This makes possible a quick floating zone growth of large single crystals without using solvent or crucibles. Together with the recent report of successful growth of ordered anion-deficient Ca$_2$Co$_2$O$_5$ in flowing oxygen atmosphere of 145 atm \cite{Ca2Co2O5}, our growth of BaFe$_{12}$O$_{19}$ in 100 atm oxygen suggests that the high pressure floating zone technique will be a useful tool for new and metastable materials in single crystal form.

\subsection{Local structure distortion of TBP site}
Our neutron single crystal diffraction study confirms the OMP displacement, although in a reduced magnitude, of TBP Fe ions at 4~K. We also observed the displacement of the apical O1 toward the TBP Fe. As neutrons are better than X-rays at scattering from light atoms like oxygen, it helped resolve this feature. The details of the observed local structure change around the TBP Fe site are illustrated in Fig. \ref{localstr}. The in-plane Fe2-O3 bonds show no temperature dependence. As highlighted by the horizontal dashed lines, the O1-O1 distance is reduced when cooling from 295 K to 4 K, resulting from the displacement of the apical O1 atoms toward Fe2. Intuitively, the displacement of O1 reduces the size of TBP which can lead to the reduced OMP displacement of Fe2. However, the OMP displacement of Fe2 is reduced by half when cooling from 295 K to 4 K, which cannot be attributed to the relatively small oxygen shift only. Instead, the oxygen shift is likely an effect of the reduced OMP displacement of Fe2.

The bipyramid can be looked as two face-sharing tetrahedra. Each tetrahedron has a low potential center that tracks the Fe$^{3+}$ ion. Therefore, there exists a double potential well (see Fig. \ref{localstr}) in each bipyramid site. The hopping of the TBP Fe ion between the tetrahedron site wells leads to the OMP displacement. The OMP displacement induces local electric dipoles, and the long range dipole-dipole Coulomb interaction usually stabilizes the local dipole into a ferroelectric or antiferroelectric order. In BaFe$_{12}$O$_{19}$, quantum fluctuations suppress the long range dipole order and no electric order has been observed down to temperatures of 0.4 K. At high temperature, a larger OMP Fe$^{3+}$ displacement was observed in our study and a M{\"o}ssbauer study \cite{moss2}, which indicates an anharmonic oscillator potential as shown in Fig. \ref{localstr}. Different from the constant equilibrium position found in a harmonic oscillator potential, the equilibrium position shifts away from the center at higher energy, i.e., higher temperature, as found here. Strong anharmonic vibrations of iron ions were also indicated in polarized infrared spectroscopy measurements \cite{pinf}.

One question that remains is whether the displacement change of O1 and Fe2 upon cooling is continuous or discontinuous. The absence of any noticeable anomaly in the temperature dependence of specific heat and electrical resistivity suggests that the displacement takes place in a continuous manner. This is consistent with the M{\"o}ssbauer study showing a continuous decrease of TBP Fe2 displacement upon cooling \cite{moss2}. To further confirm this, we monitored the evolution with temperature of the integrated intensity of (0 0 22) reflection in the temperature range 4 K-300 K with single crystal neutron diffraction. The intensity of (0 0 22) peak increases also in a continuous manner upon cooling from room temperature (results not shown).

\subsection{Magnetic quantum paraelectric}
The temperature dependence of the dielectric function, especially the non-classical $T^2$ temperature dependence of the inverse dielectric function below 20~K and an anomalous upturn below 1.40~K, strongly suggests that BaFe$_{12}$O$_{19}$ is a ferrimagnetic quantum paralectric system on the verge of ferroelectricity. It is, in general, rare to find an ion that can play both magnetic and ferroelectric roles due to the empirical "d0-ness" rule in transition metal oxides\cite{matthias, hill}. The problem comes from the fact that the formation of electric dipoles usually requires hybridization between empty \textit{d}-orbitals of cations and the filled \textit{2p} oxygen orbitals, while the magnetic properties derive from partially filled \textit{d}-orbitals. Recently, perovskite manganites, AMnO$_3$ (A=Ca, Sr, Ba), were studied both theoretically and experimentally and the shift of magnetic Mn$^{4+}$ ions in MnO$_6$ octahedra was found to produce electric dipoles \cite{maganite-1,maganite-2,maganite-3,maganite-4,maganite-5}. In BaFe$_{12}$O$_{19}$, the TBP Fe ions contribute to both magnetic and local dipoles. It thus provides another materials platform which can be tuned to optimize the cross-coupling between magnetism and ferroelectricity.

A recent study of dielectric properties of Ba$_{1-x}$Sr$_x$Fe$_{12}$O$_{19}$ suggests that the local dipole moment decreases dramatically with the substitution of Ba by Sr \cite{shen}. The reduced dipole moment is accompanied by the suppressed displacement of TBP Fe ions. This signals the importance of the displacement of TBP Fe ions for the dielectric properties of hexaferrites. It also indicates that SrFe$_{12}$O$_{19}$ is further away from the antiferroelectric border, as schematically demonstrated in Fig. \ref{dgm}. With stress engineering, as suggested in Ref. \cite{tan}, it might be possible to realize room temperature multiferroics by tuning the OMP displacement.

\section{Conclusions}
High quality single crystals of BaFe$_{12}$O$_{19}$ have been grown using the floating zone technique in flowing oxygen atmosphere of 100 atm. Single crystal neutron diffraction confirmed that the displacement of TBP Fe ions decreases continuously with decreasing temperature. The displacement at 4~K is nonzero but about half of that at room temperature. The reduced displacement of TBP Fe ions is accompanied by the oxygen displacement toward the TBP Fe ions. The temperature dependence of the dielectric function shows a non-classical $T^2$ temperature dependence of the inverse dielectric function below 20~K and an anomalous upturn below 1.40~K extending to the millikelvin range. These results provide strong evidence that BaFe$_{12}$O$_{19}$ is a ferrimagnetic quantum paraelectric compound on the border of ferroelectricity. As the TBP Fe ions contribute to both magnetism and electric dipoles, BaFe$_{12}$O$_{19}$ offers an ideal materials platform for tuning the coupling of magnetism and ferroelectricity.

\section{Acknowledgments}
Work at ORNL was supported by the US Department of Energy, Office of Science, Basic Energy Sciences, Materials Sciences and Engineering Division (BCS and JQY), and Scientific User Facilities Division (HBC). Magnetization measurements (MAM) were supported by US DOE, Energy Efficiency and Renewable Energy, Vehicle Technologies Office, Propulsion Materials Program. DGM acknowledges support from the Gordon and Betty Moore Foundation’s EPiQS Initiative through Grant GBMF4416. The work at NHMFL is supported by NSF-DMR-1157490, US DOE, and the State of Florida.

\end{document}